\documentclass[prd,aps,reprint,superscriptaddress,nofootinbib]{revtex4-2}

\pdfoutput=1

\usepackage[colorlinks=true
,urlcolor=blue
,anchorcolor=blue
,citecolor=blue
,filecolor=blue
,linkcolor=blue
,menucolor=blue
,linktocpage=true
,pdfproducer=medialab
,pdfa=true
]{hyperref}
\usepackage[T1]{fontenc}
\usepackage{fontawesome}
\usepackage{feynmf}
\usepackage{graphicx}
\usepackage{enumitem}
\usepackage{latexsym}
\usepackage{amsfonts}
\usepackage{amssymb}
\usepackage{mathrsfs}
\usepackage{color}
\usepackage{amsmath}
\usepackage[capitalise]{cleveref}
\usepackage{slashed}
\usepackage{dcolumn}
\usepackage{verbatim}
\usepackage{comment}
\usepackage{float}
\usepackage{multirow}
\usepackage{xspace}
\usepackage[dvipsnames]{xcolor}
\usepackage[normalem]{ulem}

\newcommand{\GeV}{~\text{GeV}}

\definecolor{tgreen}{cmyk}{0.65, 0, 1.0, 0.1}

\def\triumf{TRIUMF, 4004 Wesbrook Mall, Vancouver, BC V6T 2A3, Canada}
\def\sfu{Department of Physics, Simon Fraser University, Burnaby, BC V5A 1S6, Canada}
\def\uvic{Department of Physics and Astronomy, University of Victoria, Victoria, BC V8P 5C2, Canada}

\begin{document}

\title{Neutrino Electroweak Couplings and the Neutrino Fog at Belle II}

\author{Carlos Henrique de Lima}
\email{cdelima@triumf.ca}
\affiliation{\triumf}

\author{David McKeen}
\email{mckeen@triumf.ca}
\affiliation{\triumf}

\author{Afif Omar}
\email{aomar@triumf.ca}
\affiliation{\triumf}
\affiliation{\uvic}

\author{Douglas Tuckler}
\email{dtuckler@triumf.ca}
\affiliation{\triumf}
\affiliation{\sfu}

\begin{abstract}
The mono-photon process $e^+e^- \to \gamma+\mathrm{invisible}$ at Belle II probes both invisible new states and the electroweak couplings of neutrinos. In this work, we study the process $e^+e^- \to \gamma\nu\bar{\nu}$, finding that Belle II with the full expected dataset can determine the neutrino-sensitive effective weak mixing angle to a 3.6$\%$ relative statistical precision, competitive with existing neutrino probes at other energy scales. Polarized beams, as proposed for the Chiral Belle upgrade, separate the neutral- and charged-current contributions and provide independent access to each, giving a handle on the chiral structure of the neutrino gauge interactions. The same process is an irreducible background to dark boson searches, producing a neutrino fog once the luminosity exceeds 1 ab$^{-1}$.
\end{abstract}

\maketitle

\section{Introduction} 
\label{sec:intro}

The mono-photon process $e^+e^- \to \gamma + \text{invisible}$ is a powerful probe of invisible new states at colliders. At Belle II, it has been used extensively to search for dark bosons and dark matter in association with a photon~\cite{Belle-II:2018jsg,Essig:2013vha,Graham:2021ggy,deLima:2025pzd}. As Belle II luminosity approaches its anticipated total of $50~\text{ab}^{-1}$~\cite{Belle-II:2010dht}, previously subdominant backgrounds will play an increasingly important role.

One such background is the Standard Model (SM) reaction $e^+e^- \to \gamma\nu\bar{\nu}$. This process is sensitive to both the neutral and charged currents of the neutrino sector~\cite{Ma:1978zm,Dicus:1979xma}. Its final state is identical to that of an invisibly decaying dark boson: a single hard photon and missing energy. Unlike the QED backgrounds from $e^+e^- \to \gamma\gamma(\gamma)$ and $e^+e^- \to e^+e^-\gamma(\gamma)$, the neutrino background cannot be removed within the signal region by kinematic cuts. Consequently, as luminosity accumulates, this irreducible background introduces a ``neutrino fog'' in analogy with that in direct dark matter detection~\cite{Monroe:2007xp,Vergados:2008jp,Strigari:2009bq,Gutlein:2010tq,Gaspert:2021gyj,Carew:2023qrj}.

At the same time, the process $e^+e^- \to \gamma\nu\bar{\nu}$ is an interesting electroweak (EW) physics target in its own right. In the SM, the overall rate is fixed once the weak mixing angle is specified. Measuring this rate at $\sqrt{s} \approx 10.58~\text{GeV}$ provides a direct determination of the neutrino's effective couplings to the $Z$ and $W$ bosons at a momentum scale below the $Z$ pole. This connects directly to the Chiral Belle program~\cite{Roney:2019til,USBelleIIGroup:2022qro}, one of whose primary goals is to measure the effective weak mixing angle for each accessible fundamental fermion with extreme precision. Interestingly, measuring neutrino interactions is significantly harder, and there are outstanding puzzles in their extraction close to the energy scale probed at Belle II, as shown by the anomalous NuTeV result~\cite{NuTeV:2001whx}.

In a previous work~\cite{deLima:2025pzd}, we characterized the reach of the mono-photon channel for dark bosons at (Chiral) Belle~II. In this work, we turn to the irreducible Standard Model process $e^+e^- \to \gamma\nu\bar{\nu}$ that populates the identical final state, and show that it is both a precision electroweak target and the ultimate limitation on invisible dark boson searches.

\begin{figure}[b!]
    \centering
    \hspace*{-0.05\linewidth}
    \includegraphics[width =1.0\linewidth]{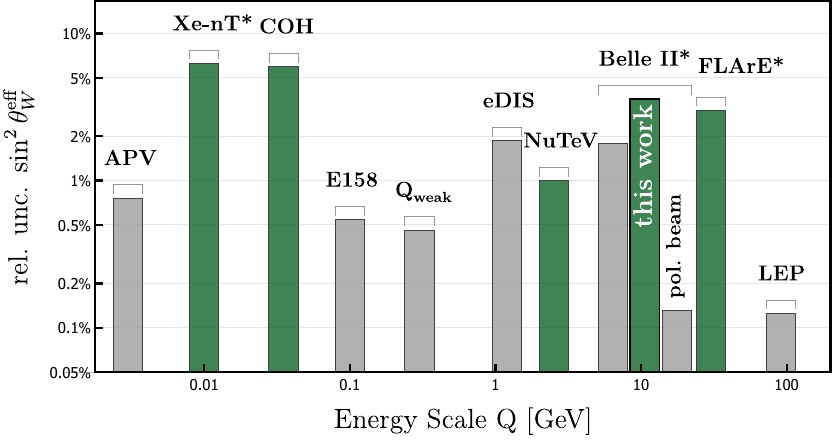}
    \caption{Relative uncertainty on the effective weak mixing angle at different energy scales~\cite{Dzuba:2012kx,SLACE158:2005uay,Qweak:2018tjf,Wang:2014guo,ALEPH:2010aa,USBelleIIGroup:2022qro,Grussbach2022WeinbergAngleBelleII,Maity:2024aji,XENON:2026ydt,MammenAbraham:2023psg,COHERENT:2021xmm,COHERENT:2020iec,NuTeV:2001whx,L3:1991tpr,OPAL:2000puu}. In green, we show the measurements that are sensitive to neutrino physics while gray bars show those that do not involve neutrinos. Asterisks denote projections. The Belle II projection in this work assumes the full unpolarized 50~ab$^{-1}$ dataset. We also show the reach of the Chiral Belle upgrade for the visible-fermion weak mixing angle extraction~\cite{USBelleIIGroup:2022qro}, assuming 40 ab$^{-1}$ of data.  }
    \label{fig:lim_effW}
\end{figure}

In this letter, we examine the mono-photon channel at (Chiral) Belle II with two complementary goals. First, we study the physics potential of measuring this process as a probe of the neutrino interactions with the $W$ and $Z$ at Belle II energies. Interpreting the measurement as an extraction of the effective weak mixing angle of the neutrinos, we show the expected precision with the full unpolarized dataset in Fig.~\ref{fig:lim_effW} together with other experimental extractions. We then consider the additional analyzing power of including polarized beams~\cite{USBelleIIGroup:2022qro,Roney:2019til}, where the individual contributions from the $Z$ and $W$ bosons can be disentangled. Second, we characterize the ``neutrino fog'' that arises in dark photon searches once the luminosity is sufficiently high.

We find that the full unpolarized dataset can determine the effective neutrino weak mixing angle to the few-percent level, competitive with existing neutrino-sensitive extractions. At the same time, the irreducible $\nu\bar\nu$ contribution begins to limit dark boson searches once the luminosity exceeds $1\,\mathrm{ab}^{-1}$. Including beam polarization reduces the systematics of the unpolarized analysis and provides additional information to differentiate contributions from $W$ and $Z$ processes.

\section{Neutrino Production in the Mono-Photon Channel}
\label{sec:signal}

The SM process $e^+e^- \to \gamma\nu\bar{\nu}$ arises from two classes of diagrams. The neutral-current diagram involves $Z$-boson exchange in the $s$-channel and contributes to the production of all three neutrino flavors ($\nu_e$, $\nu_\mu$, $\nu_\tau$). The charged-current diagram involves $W^\pm$ exchange and contributes only to $\nu_e\bar\nu_e$ production. The photon is attached to the $e^\pm$ lines in both cases; the contribution from attaching it to the $W$ is suppressed by $G_F \times s$ and can be neglected at Belle II energies. 

In the center-of-mass frame, we denote the photon energy and angle with respect to the electron beam direction as $E_\gamma$ and $\theta$, respectively, and define a convenient reference cross section at $\sqrt s=10.58~\rm GeV$,
\begin{equation}
\begin{aligned}
&\frac{d\bar\sigma}{dE_\gamma d\cos\theta} 
\approx \frac{4.6~\text{fb}}{E_\gamma}
\!\left(1 - \frac{2E_\gamma}{\sqrt s}\right)
\\
&\quad \times
\frac{\!\left(1 - E_\gamma/\sqrt s\,\right)^{\!2} + E_\gamma^2\cos^2\theta/s}
{1-\cos^2\theta}\, .
\end{aligned}
\label{eq:dsigmabar}
\end{equation}
In terms of this reference cross section, the cross sections for fully polarized left- or right-handed $e^-$ beams are
\begin{align}
\frac{d\sigma^{\nu\bar\nu}_L}{dE_\gamma d\cos\theta}
&= \left(\tfrac{1}{2} - \tfrac{2}{3} s_{\theta_W}^2 + 2s_{\theta_W}^4\right) \frac{d\bar\sigma}{dE_\gamma d\cos\theta},\\
\frac{d\sigma^{\nu\bar\nu}_R}{dE_\gamma d\cos\theta}
&= 2s_{\theta_W}^4\, \frac{d\bar\sigma}{dE_\gamma d\cos\theta}\, ,
\label{eq:sigR_tot}
\end{align}
where $s_{\theta_W}^2 \equiv \sin^2 \theta_W$, with $\theta_W$ the weak mixing angle. A right-handed electron beam couples to the neutrinos only through the $Z$, so $\sigma_R$ isolates the neutral current. The $V-A$ structure of the interactions induces a polarization dependence of this process that is not present in the (parity conserving) QED backgrounds from $e^+ e^- \to \gamma\gamma(\gamma)$ and $e^+ e^- \to e^+e^-\gamma(\gamma)$~\cite{deLima:2025pzd}~\footnote{While leading order QED has no dependence on the initial beam polarization, the outgoing photon is polarized depending on the process and this information could be recovered in the Belle II detector~\cite{deLima:2026jbb}.}. This can be further exploited with polarized beams to lift degeneracies among the possible Lorentz structures of the neutrino interactions, which an unpolarized rate measurement cannot separate. Note that the polarization dependence is only in the overall normalization and not in the kinematic distributions. Additionally, the photon c-o-m energy ranges from $0$ to $\sqrt s/2=5.29~\rm GeV$, unlike the on-shell dark boson case where it is monochromatic.

\subsection*{Parametrization of Deviations from the SM}\label{subsec:param}

We can interpret measurements of this process in terms of an effective weak mixing angle. We express the polarized cross sections in terms of this effective angle as
\begin{align}
&\sigma(s_{\theta_W^{\rm{eff}}})_P = f_P\left(s_{\theta_W^{\rm{eff}}}^2\right) \sigma_P\, ,
\\
&f_P\left(x\right)=\frac{0.89-1.19x+7.11x^2+0.89(1-1.33x)P}{1+0.61P}\, ,\nonumber
\end{align}
where the polarization $P=L^--R^-$ with $L^-$ and $R^-$ the fraction of left- and right-handed electrons in the beam. This is normalized such that $f_P\left(s_{\theta_W}^2\right)=1$ when the weak mixing angle takes its SM value of $s_{\theta_W}^2 = 0.2335$~\cite{ParticleDataGroup:2026aaa} in the $\overline{\rm MS}$ scheme and energy scale $Q\approx 10.58~$GeV. In what follows we will focus on nominal polarization values $P=0$, $+0.7$, and $-0.7$ (unpolarized, 70\% L, and 70\% R $e^-$ polarizations, respectively)~\cite{USBelleIIGroup:2022qro,Roney:2019til}. 

Another interpretation can be constructed in terms of the individual modifications to the $W$ and $Z$ couplings. In a relatively model-independent way, we can parametrize the cross sections in terms of two signal-strength parameters 
\begin{equation}
\mu_Z^{\phantom{\dagger}}  \equiv \left(\frac{g_{Zee}^{\phantom{\dagger}}g_{Z\nu\nu}^{\phantom{\dagger}}}{g_{Zee}^{\rm SM}g_{Z\nu\nu}^{\rm SM}}\right)^2, \qquad
\mu_W^{\phantom{\dagger}} \equiv \left(\frac{g_{We\nu_e}^{\phantom{\dagger}}}{g_{We\nu_e}^{\rm SM}}\right)^4.
\label{eq:muZW}
\end{equation}
In this parametrization, we write
\begin{equation}
\sigma(\mu_W,\mu_Z)_P = f_P\left(\mu_W,\mu_Z\right) \sigma_P\, ,
\end{equation}
with
\begin{equation}
\begin{aligned}
    &f_P\left(\mu_W,\mu_Z\right) = \Big[(1.19\mu_W-0.63\sqrt{\mu_W\mu_Z})(1+P)
    \\
    &\qquad\qquad +0.45\mu_Z(1+0.13P)\Big]\left(1+0.61P\right)^{-1}
\end{aligned}
\end{equation}
which is normalized to unity at the SM point $\mu_Z = \mu_W = 1$.

The parametrization in terms of modifications to the charged and neutral currents allows a finer separation between flavor-specific models. The charged current is the dominant contribution to this process, and it is sensitive exclusively to modifications of the electron-flavor neutrinos. This measurement can also be parameterized in terms of the number of neutrino species, where the neutral-current contribution has a factor of $N_\nu/3$ in the Standard Model. Explicitly, the modification factor is
\begin{align}
    f_P(N_\nu) &= \frac{0.55(1+P)+0.45(1+0.13P)(N_\nu/3)}{\left(1+0.61P\right)} \, .
\end{align}

 \begin{figure}[b!]
    \centering
    \hspace*{-0.05\linewidth}
    \includegraphics[width = 0.98\linewidth]{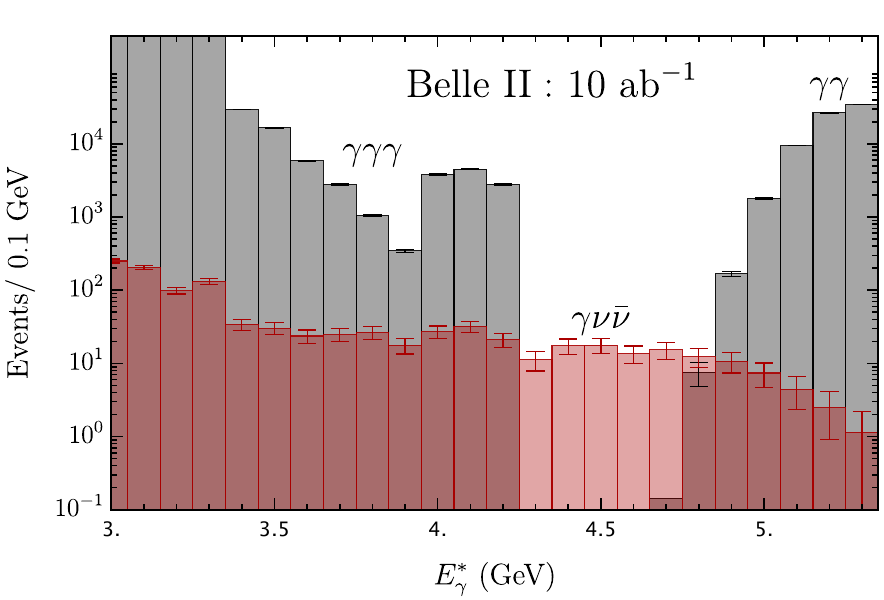}
    \caption{Detector-level histogram of the photon energy in the center-of-mass frame, with the $\theta_{\rm{lab}}$ cut explained in the main text. The dominant contributions from $\gamma\gamma\gamma$ and $\gamma \gamma$ are shown in their respective photon energy regions (in gray). The background-free region is where the reach of a dark photon search is maximized, and also where the neutrinos are present with an irreducible contribution (in red).}
    \label{fig:cut}
\end{figure}

Measuring this process thus provides a window into neutrino interactions at Belle II energies. Interestingly, the NuTeV experiment~\cite{NuTeV:2001whx} measured the effective neutrino weak mixing angle in tension with the SM prediction. The extraction of  $s^{2}_{\theta_W}$  comes from the ratio of neutral- to charged-current cross sections in $\nu$-nucleon deep inelastic scattering at a scale $Q\sim{\rm GeV}$. The NuTeV result was approximately three standard deviations away from the global EW fit value based on measurements at scales above and below that of NuTeV. While nuclear effects, isospin violation, and strange-sea asymmetries~\cite{Davidson:2001ji,Miller:2002xh} could account for some of the NuTeV discrepancy, better understanding of neutrino interactions at comparable $Q$, such as the process we study, would be very desirable.

Deviations in measurements of neutrino-sensitive processes could be generated by new neutrinophilic interactions. New forces that couple predominantly to SM neutrinos appear in several well-motivated extensions of the SM, e.g. in neutrino mass models~\cite{Chikashige:1980ui,Gelmini:1980re,Schechter:1981cv,Barger:1981vd,deLima:2022dht} and sterile neutrino dark matter models~\cite{DeGouvea:2019wpf,Kelly:2020pcy,Kelly:2020aks,An:2023mkf}. Neutrino self-interactions would not only modify this process but also generate other unique signatures at high-energy colliders~\cite{deLima:2024ohf,deGouvea:2019qaz,Dev:2021axj,Agashe:2024owh,Foroughi-Abari:2026ofc}, low-energy experiments~\cite{Pasquini:2015fjv,Berryman:2018ogk,Kelly:2019wow,Brdar:2020nbj,Deppisch:2020sqh,Zhang:2024meg,deLima:2026fsy,Altmannshofer:2026opc}, astrophysical processes~\cite{Kolb:1987qy,Ng:2014pca,Shoemaker:2015qul,Heurtier:2016otg,Das:2017iuj,Kelly:2018tyg,Bustamante:2020mep,Esteban:2021tub,Chang:2022aas,Chen:2022kal,Fiorillo:2022cdq,Fiorillo:2023ytr,Fiorillo:2023cas,Telalovic:2024cot,Akita:2022etk,Akita:2023iwq,Parashari:2026dxo}, and cosmology~\cite{Cyr-Racine:2013jua,Archidiacono:2013dua,Huang:2017egl,Escudero:2019gvw,Barenboim:2019tux,Blinov:2019gcj,Lyu:2020lps,Camarena:2024daj}. 

A measurement of neutrinos at Belle II probes the weak mixing angle at $Q \sim \sqrt s\simeq  10~\text{GeV}$, which sits between the atomic parity violation regime and the $Z$-pole measurements at LEP. More importantly, this measurement is sensitive to neutrino physics, whereas most other measurements use asymmetries involving visible leptons in the final state. The running of $s_{\theta_W}^2$ with the energy scale is predicted precisely within the SM~\cite{Erler:2004in,Erler:2017knj}. Any particle-specific deviation of the expected $s_{\theta_W}^2$ could indicate new physics that couples preferentially to the given particle. Because the effective weak mixing angle can be modified for specific EW sectors, it is ideal to measure the neutrino interactions directly as we propose in the next section.

\section{Measuring Neutrinos at (Chiral) Belle II}
\label{sec:neutrino}

To extract sensitivity to the $e^+e^- \to \gamma\nu\bar{\nu}$ process and its dependence on $(\mu_W^{\phantom{\dagger}},\mu_Z^{\phantom{\dagger}})$, we perform a shape analysis in the photon center of mass energy, $E_\gamma^*$, and lab-frame angle, $\theta_{\rm{lab}}$. The dominant backgrounds are the $e^+e^- \gamma(\gamma)$ and $\gamma\gamma(\gamma)$ final states, which contribute when one or two of the final-state particles fall outside the detector acceptance or are otherwise not reconstructed; for the di-photon background, this proceeds through photon detection inefficiency. We describe these backgrounds in more detail in~\cite{deLima:2025pzd}. We generate 90 million Monte Carlo events for each background and signal using \texttt{MadGraph5\_aMC@NLO}~\cite{Alwall:2014hca}. 

We reconstruct photons in the electromagnetic calorimeter only in the region where efficient photon reconstruction is available, $18.5^\circ < \theta < 139.2^\circ$~\cite{Belle-II:2018jsg}, and with energies $E_\gamma > 3\GeV$. We consider charged particles to be missed by the detector if they carry transverse momentum less than $p_T^{\rm min}=0.2~\rm GeV$, or do not leave a track (i.e., outside of $17^\circ < \theta_{e^\pm} ^{\text{lab}}<150^\circ$). We do not consider cosmic-ray contamination, which could affect the reach of these searches if not properly treated. We smear the photon energy with a Gaussian to simulate the Belle II energy resolution. In the region of interest, the energy resolution is approximately constant, $\sigma_E/E = 0.0225$~\cite{Adachi:2018qme}. For photons within the acceptance, we assign a detection inefficiency of $10^{-5}$~\footnote{The current tracker plus ECL design has a radiation length of $\sim 16.5X_0$~\cite{Belle-II:2018jsg,Belle-II:2010dht}, which means a probability of $\sim 3 \times 10^{-6}$ that the photon does not interact. Those missed photons could be reconstructed with lower efficiency in the KLM. The sensitivity of this study does not heavily depend on this, and while it is possible for future upgrades to reduce the probability of missing a photon below $10^{-6}$, we consider here a conservative $10^{-5}$ inefficiency. }, which is the value required to reproduce the background levels of the existing Belle~II mono-photon analysis~\cite{Belle-II:2018jsg}.  This inefficiency probability considers only sources of inefficiency that do not come from the detector geometry and gaps in the detector, which are otherwise included in the analysis.

\begin{figure}[t!]
    \centering
    \hspace*{-0.08\linewidth}
    \includegraphics[width =0.98\linewidth]{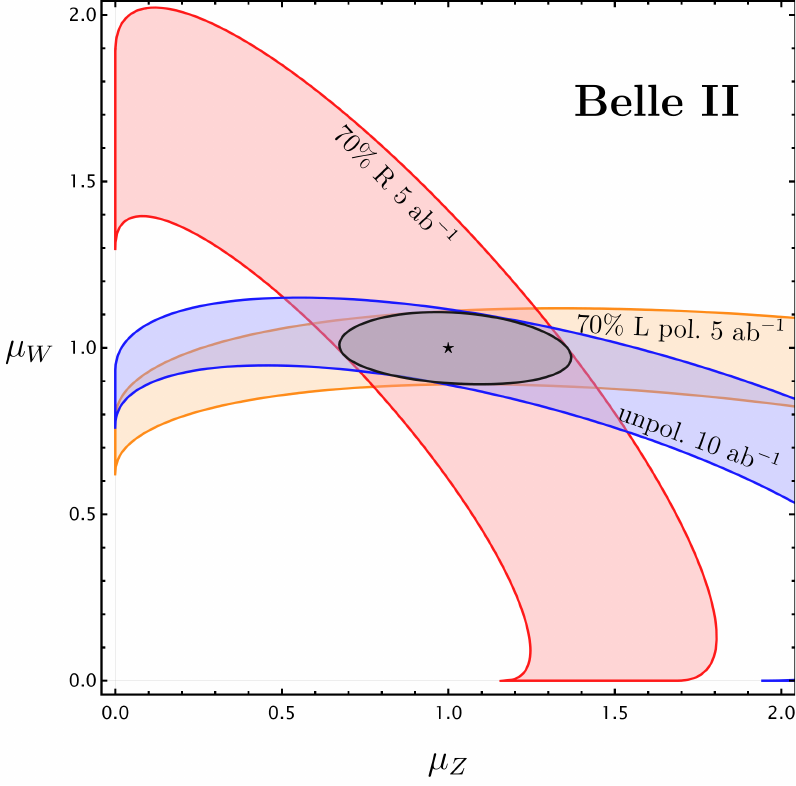}
    \caption{Projected 95$\%$ sensitivity in the $(\mu_W^{\phantom{\dagger}},\mu_Z^{\phantom{\dagger}})$ plane from $10~\mathrm{ab}^{-1}$ at Belle II. Results are shown for unpolarized beams and for $70\%$ left- and right-polarized electron beams. In black, the combination of the left- and right-polarized runs, which reduces the degeneracy of the unpolarized extraction. }
   \label{fig:lim}
\end{figure}

The background is concentrated close to the beamline; for better visualization, it is instructive to apply a cut on the inner angles. We optimize a selection on the photon polar angle $\theta_{\rm{lab}}$ to maximize sensitivity to the signal. This is performed through a scan over angular intervals $(\theta_{\min},\theta_{\max})$ maximizing $S/\sqrt{S+B}$ based on the expected signal and background yields. The optimization is carried out independently in each energy bin. This selection is used only to produce the one-dimensional visualization of Fig.~\ref{fig:cut}; all sensitivities quoted below come from the full two-dimensional binned Poisson likelihood in $(E_\gamma,\theta)$, which retains the complete shape information.

We bin the data in two dimensions, with $0.1~\mathrm{GeV}$ energy bins and $1^\circ$ angular bins, and perform a binned Poisson likelihood analysis. In each bin $i$, the expected number of events is $N_i(\mu_W,\mu_Z) = S_i(\mu_W,\mu_Z) + B_i$, where $S_i$ and $B_i$ are the numbers of signal and background events. To highlight the near-future possibility of this measurement, we carry out the analysis for 10 ab$^{-1}$ of data, considering first just an unpolarized beam. Then, we consider the additional information that splitting the data between L and R provides, with the proposed $70\%$ polarization of the electron beam. We perform a one-parameter extraction of the precision of the measurement of this cross section and a two-parameter extraction in terms of $\mu_W$ and $\mu_Z$. The individual 95$\%$ sensitivity and the combination can be seen in Fig.~\ref{fig:lim}. Most of the sensitivity for this search comes from background-free regions in the kinematic space. For the full expected dataset of 50 ab$^{-1}$, we can estimate the sensitivity for the neutrino effective weak mixing angle shown in Fig.~\ref{fig:lim_effW}. This same measurement can be interpreted as a determination of the number of neutrino species, $N_\nu = 3.00 \pm 0.12$ with the full unpolarized dataset, competitive with similar extractions at higher energies~\cite{L3:1991tpr,OPAL:2000puu,ParticleDataGroup:2026aaa}. 

We do not include systematic uncertainties in these projections. The dominant contributions are the luminosity normalization, the photon reconstruction efficiency and energy scale, and the normalization of the QED backgrounds. The left-right cross-section ratio available with polarized running cancels the luminosity and efficiency terms to a large extent, which is the main advantage of polarization for this measurement. 

A limiting factor for this study and for general mono-photon searches is the irreducible inefficiency of the photon detector, which we modeled conservatively to be $10^{-5}$. This contaminates an otherwise QED-free region with high-energy photons from $\gamma \gamma$. Reducing this inefficiency would strengthen the limits of this work and also the standard dark-sector search as discussed in~\cite{deLima:2025pzd}.

\subsection*{Other Constraints}

To compare our results to other extractions of the weak mixing angle, in Fig.~\ref{fig:lim_effW}, we place our projection among existing extractions of the effective weak mixing angle, showing the fractional uncertainties. We distinguish between measurements that do not involve neutrinos at all in gray from those that do involve neutrinos in green. The most precise determinations rely on measurements of visible-matter asymmetries and are insensitive to neutrino-specific physics: atomic parity violation~\cite{Dzuba:2012kx}, E158~\cite{SLACE158:2005uay}, Qweak~\cite{Qweak:2018tjf}, eDIS~\cite{Wang:2014guo}, the LEP Z-pole~\cite{ALEPH:2010aa}, and Belle II with~\cite{USBelleIIGroup:2022qro} and without~\cite{Grussbach2022WeinbergAngleBelleII} the Chiral Belle upgrade. The neutrino-sensitive extractions are less precise: XENONnT~\cite{Maity:2024aji,XENON:2026ydt}, FLArE~\cite{MammenAbraham:2023psg}, COHERENT~\cite{COHERENT:2021xmm,COHERENT:2020iec}, and NuTeV~\cite{NuTeV:2001whx}. With current capabilities, Belle II would provide a competitive neutrino-sector determination at $Q\approx 10\GeV$.\footnote{Similar searches involving the same process, $e^+ e^-\to \gamma\nu\bar\nu$, were performed at LEP at energies ranging from $80$ to $210~\rm GeV$ with comparable individual relative precisions in the determination of the number of neutrino species of $\sim$3-10\%~\cite{ParticleDataGroup:2026aaa} to what Belle II could achieve despite the much smaller cross section at $10~\rm GeV$. We do not re-interpret their measurements of $N_\nu$ in terms of an effective weak mixing angle for Fig.~\ref{fig:lim_effW} because this relies on the specific kinematic regions that are different for each of their measurements.} Like the other neutrino-sensitive measurements, this could provide insight into modifications of neutrino interactions in a way that visible-particle-specific searches are blind to.

\section{Discussion}
\label{sec:conc}

As the luminosity of Belle II increases, the mono-photon channel becomes a unique environment where precision EW measurements and dark-sector searches share the same final state. We studied the reach that could be achieved at Belle II to measure weak charged- and neutral-current interactions involving neutrinos. The expected bounds on the modifications of the effective $\bar ee\bar\nu\nu$ couplings are shown in Fig.~\ref{fig:lim}; this could also be interpreted in terms of a measurement of the effective number of light neutrinos with a fractional uncertainty of 4\% which would be very competitive with direct probes at higher energies at LEP~\cite{L3:1991tpr,OPAL:2000puu,ParticleDataGroup:2026aaa} or as a measurement of the weak mixing angle to a precision of 3.6\% at $Q\sim10~\rm GeV$ which is comparable to other measurements involving neutrinos~\cite{ParticleDataGroup:2026aaa}. 

\begin{figure}[t!]
    \centering
    \hspace*{-0.05\linewidth}
    \includegraphics[width = 1\linewidth]{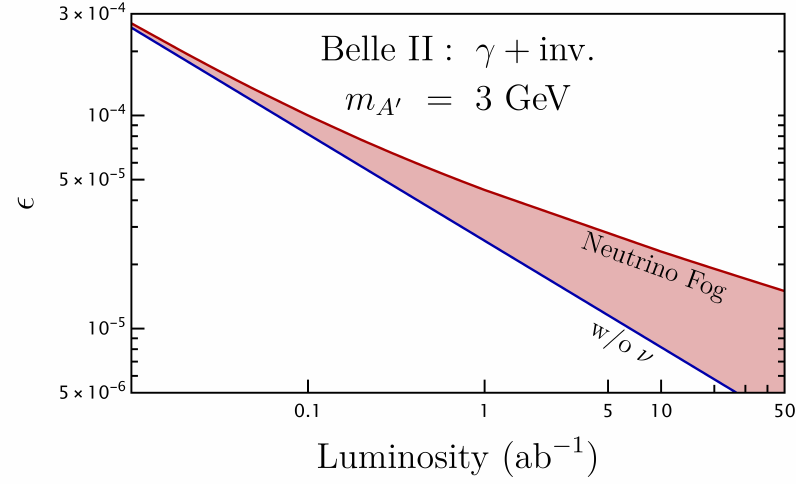}
    \caption{Sensitivity to the kinetic mixing of a $3~\mathrm{GeV}$ dark photon with and without the neutrino background. The inclusion of $e^+e^- \to \gamma\nu\bar{\nu}$ reduces the reach, illustrating the neutrino fog effect. The slope of the discovery reach changes as it enters the fog region, as seen around 1 ab$^{-1}$ of data.}
    \label{fig:fog}
\end{figure}

The process  $e^+e^- \to \gamma\nu\bar{\nu}$ also introduces a fundamental limitation on future searches for new invisible states, analogous to the neutrino fog encountered in direct dark matter experiments. The impact of this neutrino background is illustrated in Fig.~\ref{fig:fog}, where we compare the projected sensitivity to the kinetic mixing, $\epsilon$, of a $3\GeV$ dark photon, $A^\prime$, as a function of integrated luminosity with and without the SM neutrino contribution. In the absence of neutrinos, the sensitivity continues to improve with luminosity as the dominant QED backgrounds can be efficiently suppressed through kinematic selections. Once the neutrino contribution is included, however, the search becomes limited by an irreducible component with the same experimental signature as an invisibly decaying dark boson. Note that for lighter dark photon masses, the photon inefficiency of the detector produces a larger irreducible background from QED $\gamma \gamma$. At the current photon inefficiency, the neutrino fog dominates for dark photon masses roughly in the $m_{A'}\simeq 2.5\textrm{--}5\GeV$ window, and could become important below 2.5 GeV if the inefficiency is reduced.  

The change in slope around 1 ab$^{-1}$ of data in Fig.~\ref{fig:fog} comes from a change in how the reach scales with luminosity. The signal sensitivity initially scales as $\epsilon\propto\mathcal{L}^{-1/2}$, since the search is background-free. Once the neutrino contribution takes over, the background is irreducible, and the limit is set by its statistical fluctuation, improving only as  $\epsilon\propto\mathcal{L}^{-1/4}$. Note that the reach flattens if a systematic uncertainty on the neutrino normalization dominates over its statistical fluctuation. 

The same effect will appear at future lepton colliders, just as it was present at LEP. The FCC-ee~\cite{Bernardi:2022hny}, running at higher energy and luminosity, will produce the $\gamma\nu\bar\nu$ final state, so any mono-photon search for invisible states there will eventually reach its own neutrino fog. The same is true for lower energies, as in the Super Tau-Charm Facility~\cite{Ai:2025xop}, where the neutrino fog will appear at slightly higher luminosities.

\acknowledgments
We thank Michael Roney, Christopher Hearty, and Yue Zhang for helpful discussions and feedback on the draft. This work is supported by Discovery Grants from the Natural Sciences and Engineering Research Council of Canada (NSERC). TRIUMF receives federal funding via a contribution agreement with the National Research Council (NRC) of Canada.

\bibliography{ref_CBelle}

\end{document}